\documentclass{epl}
\usepackage{epsfig}
\usepackage{amssymb,amsbsy,amsmath}
\newtheorem{lemma}{Lemma}
\newtheorem{defi}{Definition}
\newtheorem{prop}{Proposition}

\newtheorem{theorem}{Theorem}
\newcommand{\xref}[1]{\protect\ref{#1}}
\newcommand{\fmref}[1]{(\protect\ref{#1})}
\newcommand{\figref}[1]{fig.~\protect\ref{#1}}
\newcommand{\op}[1]{%
    \fontdimen12\textfont3=2pt\fontdimen12\scriptfont3=1.4pt%
    \!\null\mathop{\vphantom{#1}\smash{#1}}\limits_{\sim}\null\!}

\def\bra#1{\langle \, {#1} \, | \,}
\def\ket#1{\, | \, {#1} \, \rangle}
\newcommand{\Tr}[1]{\text{Tr}\left\{#1\right\}}
\newcommand{\bi}[1]{\mathbf{#1}}
\title{Bounding and approximating parabolas for the spectrum of Heisenberg spin systems}
\shorttitle{Bounding parabolas}
\author{Heinz-J\"urgen Schmidt\inst{1} \and J\"urgen Schnack\inst{1}, \and Marshall Luban\inst{2}}
\shortauthor{H.-J.~Schmidt \and J.~Schnack \and M.~Luban}
\institute{
  \inst{1} Universit\"at Osnabr\"uck, Fachbereich Physik,
D-49069 Osnabr\"uck, Germany\\
  \inst{2} Ames Laboratory \& Department of Physics and Astronomy,
Iowa State University, Ames, Iowa 50011, USA
}

\begin{document}

\maketitle

\begin{abstract}
We prove that for a wide class of quantum spin systems with
isotropic Heisenberg coupling the energy eigenvalues which
belong to a total spin quantum number $S$ have upper and lower
bounds depending at most quadratically on $S$. The only
assumption adopted is that the mean coupling strength of any
spin w.~r.~t.~its neighbours is constant for all $N$ spins.  The
coefficients of the bounding parabolas are given in terms of
special eigenvalues of the $N\times N$ coupling matrix which are
usually easily evaluated. In addition we show that the bounding
parabolas, if properly shifted, provide very good approximations
of the true boundaries of the spectrum.  We present numerical
examples of frustrated rings, a cube, and an icosahedron.
\end{abstract}

%===================    section    =================================
\section{Introduction}

The study of finite spin systems is not only interesting in its
own right but has also important applications for understanding
so-called ``molecular magnets".  The synthesis of these systems
has undergone rapid progress in recent years building on
successes in coordination and polyoxometalate chemistry
\cite{Gat:AM94,Win:CIC99,MPP:CR98,GSC:CC00}.  Each of the
identical molecular units can contain as few as two and up to
several dozen paramagnetic ions.  The largest paramagnetic
molecule synthesized to date, the polyoxometalate
\{Mo$_{72}$Fe$_{30}$\}, contains 30 iron ions of spin $s=5/2$
\cite{MSS:ACIE99}.  It appears that in the majority of these
molecules the localized single-particle magnetic moments
(``spins") couple antiferromagnetically and the spectrum is
rather well described by the Heisenberg model
\cite{BeG90,DGP:IC93,CCF:IC95,PDK:JMMM97,WSK:IO99}.

Since the dimension of the Hilbert space for $N$ spins of spin
quantum number $s$, given by $(2s+1)^N$, grows rapidly with $N$
and $s$, the numerical evaluation of all energy eigenvalues
may be impossible even if the obvious symmetries of the
Hamiltonian are exploited \cite{BSS:JMMM00}.  Hence it is
indispensable to resort to approximation methods, either
analytical or numerical. However, it appears also desirable
to extend the body of rigorous results on spin systems, which
could serve as a basis or source of inspiration for the
development of approximate models.  For example, the
Berezin-Lieb inequalities \cite{B,L} relating spectral
properties of the quantum systems to those of their classical
counterparts provide a foundation for classical or
semi-classical treatments of spin systems.  Among the
Berezin-Lieb inequalities are the following rigorous estimates
%--------------------------------------------------------
\begin{equation}\label{0}
(s+1)^2\, E_{\text{min}}^{\text{classic}} 
\le E_{\text{min}}
\le 
s^2\, E_{\text{min}}^{\text{classic}} 
\end{equation}
%--------------------------------------------------------
for the ground state energy $E_{\text{min}}<0$ of quite general
Heisenberg model systems, where
$E_{\text{min}}^{\text{classic}}$ denotes the minimal energy of
the corresponding classical spin system described with unit
vectors.

In this article we extend the findings of Lieb and Berezin and
prove similar inequalities for the extremal values
$E_{\text{min}}(S)$ and $E_{\text{max}}(S)$ of the energy levels
belonging to a given quantum number $S$ of the total spin. It
turns out that the $S$-resolved eigenvalue spectrum is bounded
by two parabolas. The proof rests mainly on a linear
transformation of the spin observables according to the
eigenbasis of the coupling matrix ${\mathbb J}$.  The bounding
parabolas depend only on $N, s$ and three special eigenvalues
$j, j_{\text{min}}, j_{\text{max}}$ of ${\mathbb J}$.  If
$E_{\text{min}}^{\text{classic}} = N j_{\text{min}}$, which is
satisfied for all systems where
$E_{\text{min}}^{\text{classic}}$ has been determined (see
\cite{SL2}), our estimate
$s(s+1)E_{\text{min}}^{\text{classic}}$ for $S=0$
improves the lower Berezin-Lieb estimate, eq. \fmref{0}.
Hence we conjecture that also for arbitrary $S$ our bounds yield
the correct $s^2$-term of the asymptotic series of
$E_{\text{min}}(S)$ and $E_{\text{min}}(S)$ for $s\rightarrow
\infty$.

We compare the derived bounds with the actual numerically
calculated eigenvalue spectrum for various systems. As expected,
the bounds are rather close to the actual spectrum if $s$ is
large, but there is a considerable gap for small $s$.  For
special bipartite systems which we call ``homogeneous
Lieb-Mattis systems" we have derived a closer lower bound even
for small $s$. 

In addition, we provide numerical evidence that the bounding
parabolas, if properly shifted, turn out to be very good
approximations of the true boundaries of the spectrum, which is
especially useful for the approximation of low-lying
excitations.

%===================    section    =================================
\section{Definitions and simple results}

We consider spin systems with $N$ spin sites, spin quantum
number $s$ and isotropic Heisenberg coupling between all sites
$\mu$ and $\nu$ with coupling constants $J_{\mu\nu}$.  Let
$\op{s}_{\mu}^{(i)}\quad (i=1,2,3)$ denote the three components of
the spin observable $\op{\vec{s}}_\mu$ at site $\mu$ and, as
usual,
%--------------------------------------------------------
\begin{equation}\label{1}
\op{\vec{S}}\equiv\sum_\mu \op{\vec{s}}_\mu ,
\quad
\op{S}^{(i)}\equiv\sum_\mu \op{s}^{(i)}_\mu,
\end{equation}
%--------------------------------------------------------
denote the total spin vector and its various components. All linear operators
occuring in this context operate on a
$dim=(2s+1)^N$-dimensional Hilbert space ${\mathcal H}$.
The eigenspace of $\op{\vec{S}}^2$ corresponding to the eigenvalue $S(S+1)$
will be denoted by ${\mathcal H}_S$.

The Heisenberg Hamilton operator can be written as
%--------------------------------------------------------
\begin{equation}\label{2}
\op{H}
\equiv
\sum_{\mu\nu}J_{\mu\nu} \op{\vec{s}}_\mu\cdot\op{\vec{s}}_\nu
\ .
\end{equation}
%--------------------------------------------------------
Throughout this Letter the
coupling constants $J_{\mu\nu}$ are assumed to satisfy
%--------------------------------------------------------
\begin{equation}\label{3}
J_{\mu\nu}=J_{\nu\mu} \qquad J_{\mu\mu}=0,\qquad j\equiv \sum_\nu J_{\mu\nu},
\end{equation}
%--------------------------------------------------------
with $j$ being independent of $\mu$. The latter may be viewed as
a kind of weak homogeneity assumption.
The matrix of coupling constants
will be denoted by ${\mathbb J}$. Its trace vanishes due to
$J_{\mu\mu}=0$.  Being symmetrical, it has a complete set of
(ordered) eigenvalues $j_1,\ldots,j_N$.  One of
them is the row sum $j$ with
$\bi{1}\equiv\frac{1}{\sqrt{N}}(1,1,\ldots,1)$ as the
corresponding eigenvector.
Let  ${\mathbb J}^\prime$ denote the matrix ${\mathbb J}$
restricted to the subspace orthogonal to $\bi{1}$,
and $j_{\text{max}}$ ($j_{\text{min}}$) the largest (smallest)
eigenvalue of ${\mathbb J}^\prime$.
Since we made no assumptions on the
signs of the $J_{\mu\nu}$, it may happen that some of the
numbers $j,j_{\text{min}},j_{\text{max}}$ coincide. In most cases of
interest, the $N\times N$-matrix ${\mathbb J}$ can be easily
diagonalized, either analytically or numerically, in contrast to
the $dim\times dim$-dimensional Hamilton matrix. We will denote
the $\alpha$-th normalized eigenvector of ${\mathbb J}$ by
$(c_{1\alpha},\ldots,c_{N\alpha})$, i.~e.~
%--------------------------------------------------------
\begin{equation}\label{}
\sum_{\nu}J_{\mu\nu} c_{\nu\alpha} = j_\alpha c_{\mu\alpha},\quad \mu,\alpha=1,\ldots,N,
\end{equation}
%--------------------------------------------------------
and
%--------------------------------------------------------
\begin{equation}\label{3a}
\sum_\mu \overline{c_{\mu\alpha}} c_{\mu\beta}=
\delta_{\alpha\beta},  \quad \alpha,\beta=1,\ldots,N,
\end{equation}
%--------------------------------------------------------
where we also allow for the possibility to choose complex
eigenvectors.  Sums over $\alpha=1,\ldots,N$ excluding
$\alpha_j$ will be denoted by $\sum'$, where $\alpha_j$ denotes
the index (within the ordered set of all
eigenvalues) of the eigenvalue $j$ belonging to the eigenvector $\bi{1}$.

For later use we will consider a transformation of the spin
observables analogous to the transformation onto the eigenbasis
of ${\mathbb J}$ and define
%--------------------------------------------------------
\begin{defi}
$\op{\vec{T}}_\alpha\equiv \sum_\mu \overline{c_{\mu\alpha}}
\op{\vec{s}}_\mu,
\text{ and  }
\op{Q}_\alpha\equiv\op{\vec{T}}_\alpha^\dagger\cdot \op{\vec{T}}_\alpha
\ , \quad \alpha=1,\ldots,N$.
\end{defi}
%--------------------------------------------------------
The inverse transformation then yields
%--------------------------------------------------------
\begin{equation}\label{}
\op{\vec{s}}_\mu= \sum_\alpha c_{\mu\alpha}
\op{\vec{T}}_\alpha,\quad \mu=1,\ldots,N.
\end{equation}
%--------------------------------------------------------
In particular, $\op{\vec{T}}_{\alpha_j}=\op{\vec{S}}/\sqrt{N}$.
The following lemma follows directly from the definitions:
%--------------------------------------------------------
\begin{lemma}
\label{L1}
$
Ns(s+1)
=
\sum_{\mu} (\op{\vec{s}}_\mu)^2
=
\sum_\alpha \op{Q}_\alpha
=
\frac{1}{N}\op{\vec{S}}^2
+ \sum'_\alpha \op{Q}_\alpha
\ .
$
\end{lemma}
%--------------------------------------------------------

%===================    section    =================================
\section{Bounding parabolas}

Our main result is formulated in the following theorem:
\begin{theorem}\label{T1}
\begin{enumerate}
\item The following operator inequality holds:
%--------------------------------------------------------
\begin{equation}
\frac{j-j_{\text{min}}}{N} \op{\vec{S}}^2+j_{\text{min}} N s (s+1)
\le \op{H} \le
\frac{j-j_{\text{max}}}{N} \op{\vec{S}}^2+j_{\text{max}} N s
(s+1)
\ .
\end {equation}
%--------------------------------------------------------
\item Consequently, if $\ket{\varphi}$ is an arbitrary
normalized vector lying in the subspace ${\mathcal H}_S$, the
following bounds hold for the expectation value of $\op{H}$:
%--------------------------------------------------------
\begin{equation}\label{E-9}
\frac{j-j_{\text{min}}}{N} S(S+1)+j_{\text{min}} N s (s+1)
\le
\bra{\varphi}\op{H}\ket{\varphi}
\le
\frac{j-j_{\text{max}}}{N} S(S+1)+j_{\text{max}} N s (s+1)
\ .
\end {equation}
%--------------------------------------------------------
\end{enumerate}
\end{theorem}

{\bf Proof:} We rewrite the Hamiltonian in the following form and conclude
%--------------------------------------------------------
\begin{eqnarray}
\label{6}
\op{H}
& = &
\sum_{\mu\nu\alpha\beta} J_{\mu\nu} \overline{c_{\mu\alpha}}c_{\nu\beta}
\op{\vec{T}}_\alpha^\dagger\cdot \op{\vec{T}}_\beta
=   \sum_\beta j_\beta \op{Q}_\beta
=  \frac{j}{N} \op{\vec{S}}^2 +  \sum_\beta{}^\prime\; j_\beta
\op{Q}_\beta
\\ \nonumber
& \ge & \frac{j}{N} \op{\vec{S}}^2 +  j_{\text{min}} \sum_\beta{}^\prime\;
\op{Q}_\beta
\\ \nonumber
& = &
\frac{j}{N} \op{\vec{S}}^2 +j_{\text{min}}\left( Ns(s+1) -\frac{1}{N}\op{\vec{S}}^2 \right)
=
\frac{j-j_{\text{min}}}{N}\op{\vec{S}}^2 +j_{\text{min}} N s(s+1)
\ ,
\nonumber
\end {eqnarray}
%--------------------------------------------------------
using (\ref{3a}), the positivity of $\op{Q}_\beta$, and lemma \ref{L1}.
The other inequality follows
analogously.\hfill $\blacksquare$

As a check of our theorem \ref{T1} we consider the mean energy
$\overline{E}(S)$ within the subspace ${\mathcal H}_S$.  Let
$\op{P}_S$ denote the projector onto ${\mathcal H}_S$ and
define
$\overline{E}(S)\equiv\Tr{\op{H}\op{P}_S}/\Tr{\op{P}_S}$.
Then we obtain the following
%--------------------------------------------------------
\begin{prop}\label{P1}
\begin{eqnarray}\label{8}
\frac{j-j_{\text{min}}}{N} S(S+1)+j_{\text{min}} N s (s+1)
& \le &
\overline{E}(S)
=
\frac{j}{N-1} \left(S(S+1)-N s(s+1) \right)
\\
& \le &
\frac{j-j_{\text{max}}}{N} S(S+1)+j_{\text{max}} N s (s+1)
\nonumber
\ .
\end{eqnarray}
\end{prop}
%--------------------------------------------------------
Anticipating that formula (\ref{8}) for $\overline{E}(S)$ is
valid, which is proven in \cite{SL1}, this proposition of course
follows from theorem \ref{T1}, but we want to give an
independent proof of the stated inequalities for the interested
reader.

{\bf Proof:} Since $S\le Ns$ and $1\le N$ it follows that
%--------------------------------------------------------
\begin{equation}\label{9}
S(S+1)\le Ns(Ns+1)\le N^2 s(s+1).
\end{equation}
%--------------------------------------------------------
Moreover,
%--------------------------------------------------------
\begin{equation}\label{10a}
0 = \Tr{\mathbb J} = j + \sum_{\alpha}{}^\prime\;
j_\alpha \le j+ (N-1)j_{\text{max}},
\end{equation}
%--------------------------------------------------------
and, analogously,
%--------------------------------------------------------
\begin{equation}\label{10b}
0\ge j+ (N-1)j_{\text{min}}.
\end{equation}
%--------------------------------------------------------
The first inequality of the proposition is equivalent to
%--------------------------------------------------------
\begin{eqnarray}\label{11}
\left((N-1)(j-j_{\text{min}})-N j \right) S(S+1) + N^2 s(s+1)(j+(N-1) j_{\text{min}} )
& \le &
0
\end{eqnarray}
%--------------------------------------------------------
or
%--------------------------------------------------------
\begin{eqnarray}\label{12}
-\left((N-1)j_{\text{min}}+j \right)\left( S(S+1)-N^2 s(s+1)\right) &
\le & 0
\ .
\end{eqnarray}
%--------------------------------------------------------
But \fmref{12} follows directly from \fmref{9} and \fmref{10b}.
The second inequality of the proposition is proven in a
completely analogous way.
\hfill $\blacksquare$

In the classical limit $s \rightarrow \infty $ the bounds of
theorem \ref{T1} assume the values
%--------------------------------------------------------
\begin{eqnarray}
\label{13}
E_{\text{lower}}(S)
& \equiv &
\frac{j-j_{\text{min}}}{N} S^2 + j_{\text{min}} N s^2
\ ,
\\
E_{\text{upper}}(S)
& \equiv&
\frac{j-j_{\text{max}}}{N} S^2 + j_{\text{max}} N s^2
. \label{13a}
\end{eqnarray}
%--------------------------------------------------------
It is plausible and can be shown analogously to the proof of
theorem \ref{T1} \cite{SL2} that the corresponding classical
Hamilton function assumes only values between these bounds. The
classical parabolas \fmref{13}, \fmref{13a} intersect in the
point $S=Ns, E=j N s^2$ which corresponds to the (quantum or
classical) ferromagnetic ground state.  For special systems with
nearest neighbour coupling $J$, including the $N$-polygon
(ring), the cube, and the octahedron, it can be shown that the
bounds $E_{\text{lower}}(S)$ and $E_{\text{upper}}(S)$ are
assumed for some classical states \cite{SL2}.

For a special class of bipartite spin systems we can extend
theorem \xref{T1} and obtain an improved lower bounding
parabola.
\begin{defi}\label{D2}
Let the set of spin sites be divided into two disjoint subsets
${\cal A}$ and ${\cal B}$ such that the coupling constants
within ${\cal A}$ or ${\cal B}$ are $\le 0$, but $\ge 0$ between
${\cal A}$ and ${\cal B}$. Moreover, we assume that the partial
row sum $j^+$ of positive entries of ${\mathbb J}$ will be
constant for all rows.
Hence the same holds for $j^-\equiv j - j^+$. Further we
assume that ${\mathbb J}$ is irreducible, i.~e.~that the spin
system cannot be decomposed into unconnected parts.

Then the spin system will be called an HLM-system (``homogeneous
Lieb-Mattis system", see \cite{Mar:PRS55,LSM:AP61,LiM:JMP62}).
Further, we will denote by $\delta$ the (positive) difference
between the second smallest eigenvalue of ${\mathbb J}{}^\prime$
and $j_{\text{min}}$.
\end{defi}

It follows that for HLM-systems $|{\cal A}| = |{\cal B}| = N/2$.
Examples are even rings or cubes with suitable coupling.
%--------------------------------------------------------
\begin{theorem}\label{T2}
For HLM systems the following operator
inequality holds:
%--------------------------------------------------------
\begin{equation}
\frac{j-j_{\text{min}}}{N} \op{\vec{S}}^2+j_{\text{min}} N s (s+1)
+
\delta (N-2) s
\le \op{H}
\ .
\end {equation}
%--------------------------------------------------------
\end{theorem}
{\bf Proof}: Due to the HLM-assumption, ${\mathbb J}$ has
an eigenvector $\frac{1}{\sqrt{N}}(1,\ldots,1,-1,\ldots,-1)$
with eigenvalue $j^- - j^+$, if the spin sites are suitably permuted.
By the theorem of Ger\v{s}gorin (c.f.~\cite{PL}, 7.2), $j^- - j^+$ will be
the smallest eigenvalue $j_{\text{min}}$ and non-degenerate
(theorem of Perron-Frobenius, c.f.~\cite{PL}, 9.2).
It follows that
%--------------------------------------------------------
\begin{equation}\label{18a}
\op{Q}_{\text{min}}
=
\op{\vec{T}}_{\text{min}}^\dagger\cdot \op{\vec{T}}_{\text{min}}
=
\frac{1}{N}(\op{\vec{S}}_{\cal A}-\op{\vec{S}}_{\cal B})^2
=
\frac{1}{N} \left( 2(\op{\vec{S}}_{\cal A}^2 +\op{\vec{S}}_{\cal B}^2)
- \op{\vec{S}}^2\right) ,
\end{equation}
%--------------------------------------------------------
where $\op{\vec{S}}_{\cal A}\equiv\sum_{\mu\in{\cal A}}\op{\vec{S}}_\mu$
and $\op{\vec{S}}_{\cal B}\equiv\sum_{\mu\in{\cal B}}\op{\vec{S}}_\mu$.
The largest eigenvalue of $ \op{Q}_{\text{min}}$ within the subspace ${\cal H}_S$
is assumed for the maximal value of the spin quantum numbers
$S_{\cal A} = S_{\cal B} = N s/2$, hence
%--------------------------------------------------------
\begin{equation}\label{18b}
\op{P}_S  \op{Q}_{\text{min}} \op{P}_S
\le
\frac{1}{N}\left( 4 \frac{Ns}{2} (\frac{Ns}{2}+1)-S(S+1)\right)
=
Ns^2+2s-\frac{1}{N}S(S+1).
\end{equation}
%--------------------------------------------------------
Thus we may improve the inequality (\ref{9}) by inserting the following term
(suppressing the $\op{P}_S$)
%--------------------------------------------------------
\begin{eqnarray}\label{18c}
\sum_{\alpha}{}^\prime\;(j_\alpha-j_{\text{min}})\op{Q}_\alpha
& \ge &
\delta \left(  \sum_{\alpha}{}^\prime\; \op{Q}_\alpha - \op{Q}_{\text{min}}\right) \\
&=&
\delta \left( Ns(s+1)-\frac{1}{N}S(S+1)- \op{Q}_{\text{min}}\right) \\
& \ge & \delta (N-2)s.
\end{eqnarray}
%--------------------------------------------------------
The last inequality follows from (\ref{18b}). This completes the proof.
\hfill $\blacksquare$

It may very well be that a derivation of an improved inequality
along this line is also possible for other than bipartite
systems. Note that we have not yet used symmetry assumptions up
to weak homogeneity \fmref{3}. The key point is to obtain better
estimates for $\op{Q}_\alpha$ than those already following from
\fmref{E-9}.

%===================    section    =================================
\section{Examples}

%===================    figure   =================================
\begin{figure}[ht!]
\begin{center}
%\onefigure{fig-1.eps}
\epsfig{file=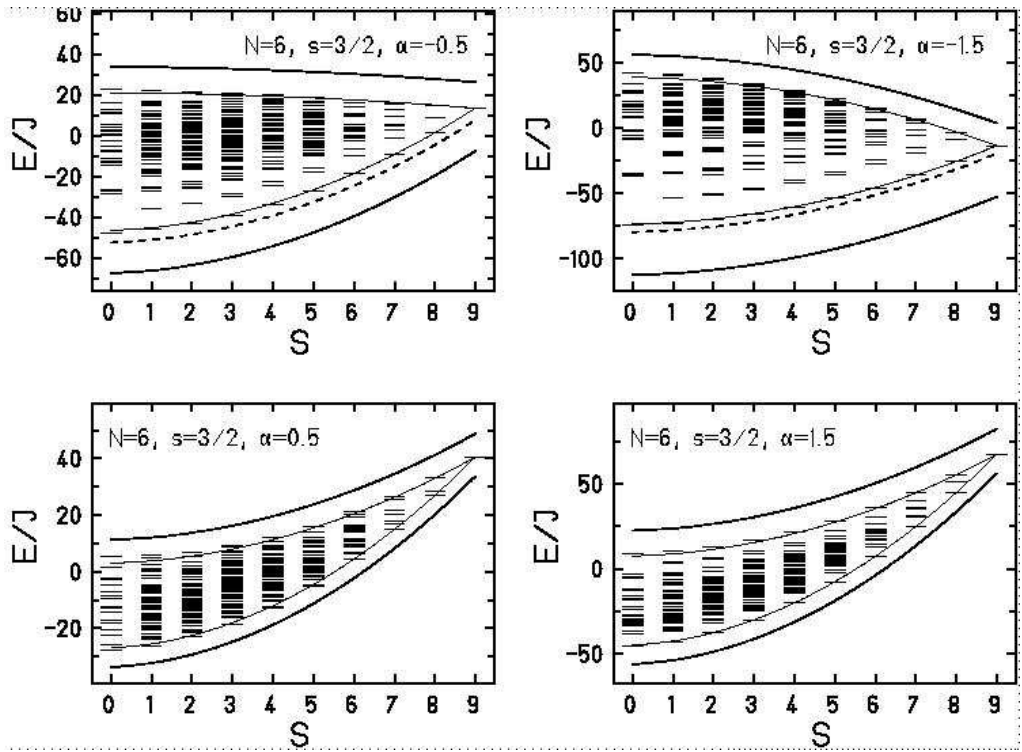,width=135mm}
\caption{Energy spectrum of Heisenberg rings with
antiferromagnetic exchange between nearest neighbours and a 
competing next-nearest-neighbour interaction. Bounding parabolas
according to theorem 
\xref{T1} are always depicted by thick solid curves, those
according to theorem \xref{T2} by dashed curves. The thin curves
depict the approximating parabolas.}
\label{F-1}
\end{center}
\end{figure}
%===================    figure   =================================
In the following figures \xref{F-1}, \xref{F-2}, and \xref{F-3}
we present examples for the bounding parabolas.
The energy spectra of \figref{F-1} have been calculated for the
Hamilton operator of spin rings with nearest neighbour and
next-nearest neighbour interaction
%--------------------------------------------------------
\begin{eqnarray}
\label{E-2}
H
&=&
2\,J\,
\left(
\sum_{\mu=1}^N\;
\op{\vec{s}}_\mu \cdot \op{\vec{s}}_{\mu+1}
+
\alpha\;
\sum_{\mu=1}^N\;
\op{\vec{s}}_\mu \cdot \op{\vec{s}}_{\mu+2}
\right)
\ ,
\end{eqnarray}
%--------------------------------------------------------
where the sum is understood $\text{mod }N$. In \figref{F-1} we
display spectra for a ring with $N=6$ and $s=3/2$ for various
ratios $\alpha$ of the two coupling constants.

%===================    figure   =================================
\begin{figure}[ht!]
\begin{center}
\epsfig{file=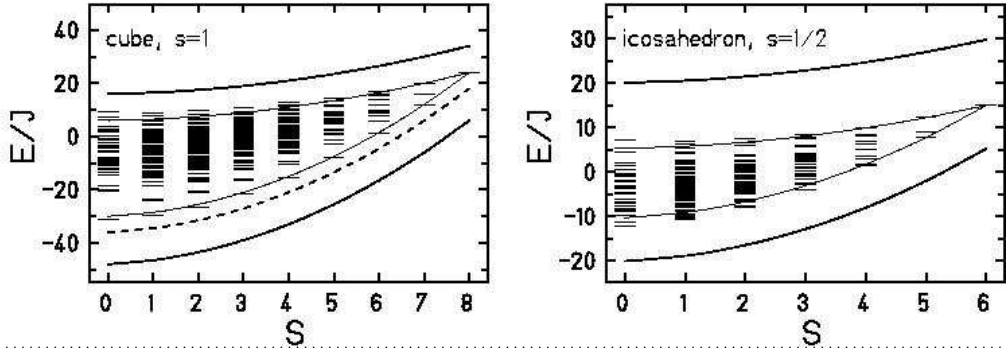,width=135mm}
\caption{Energy spectrum of an antiferromagnetically coupled
Heisenberg cube 
(l.h.s.; $j=3, j_{\text{min}}=-3,j_{\text{max}}=1$) 
and of an antiferromagnetically
coupled icosahedron 
(r.h.s.; $j=5, j_{\text{min}}=-\sqrt{5}, j_{\text{max}}=\sqrt{5}$).}
\label{F-2}
\end{center}
\end{figure}
%===================    figure   =================================
We also consider the Hamiltonian of a system of spins occupying
the vertices of a cube as well as of an icosahedron with nearest
neighbours interaction of constant strength $J$, see \figref{F-2}.

In all examples the bounding parabolas provide rigorous bounds,
but unfortunately not very narrow ones, especially for small $s$.
Only in the cases of bipartite systems the lower bound is
impressively close to the boundary of the true spectrum, even if
$s$ is small.

%===================    section    =================================
\section{Approximating parabolas}

From the figures it appears that the extremal values
$E_{\text{min}}(S)$ and $E_{\text{max}}(S)$ of the exact energy
spectrum also lie on approximate parabolas. This is not a novel
observation.  It has been noted on several occasions, that for
Heisenberg rings with antiferromagnetic nearest-neighbour
interaction and with an even number $N$ of sites the set of
minimal energies $E_{\text{min}}(S)$ forms a rotational band
\cite{CCF:CEJ96,ACC:ICA00}.
%===================    figure   =================================
%\begin{figure}[ht!]
\begin{figure}[tttt]
\begin{center}
\epsfig{file=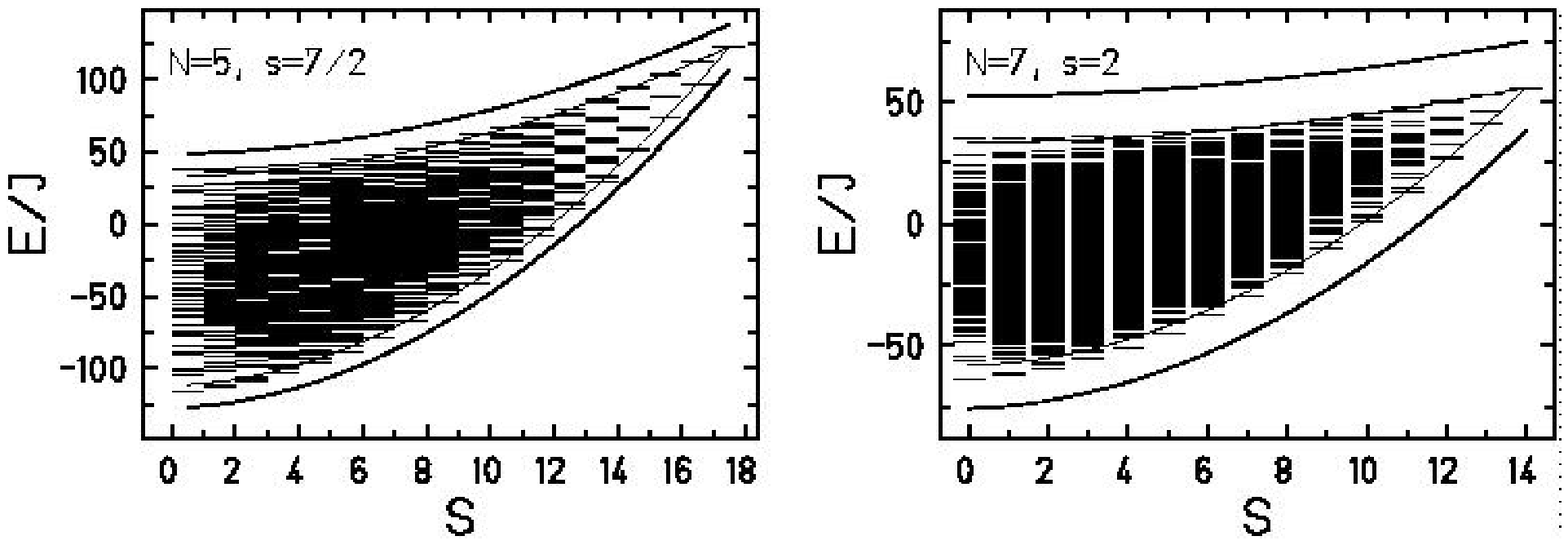,width=135mm}
\caption{Energy spectrum of antiferromagnetically coupled
Heisenberg rings with $N=5$, $s=7/2$ 
(l.h.s.; $j=2, j_{\text{min}}=-(1+\sqrt{5})/2, j_{\text{max}}=-(1-\sqrt{5})/2$) 
and $N=7$, $s=2$ 
(r.h.s.; $j=2, 
j_{\text{min}}=-1/3-\sqrt{7}/3 \cos(\phi)-\sqrt{7/3}\sin(\phi)=-1.80194, 
j_{\text{max}}=-1/3+2\sqrt{7}/3 \cos(\phi)=1.24698; 
\phi=1/3\, \text{arctan}(3\sqrt{3})$).}
\label{F-3}
\end{center}
\end{figure}
%===================    figure   =================================
Further examples, including rings for both even and odd $N$,
tetrahedra, cubes, octahedra, icosahedra, triangular prisms, and
axially truncated icosahedra, suggest that this ``rotational
band structure hypothesis" holds for general Heisenberg spin
systems \cite{ScL:PRB}.  The parabolas of our theorem \ref{T1}
have a similar curvature but a different shift constant compared
with the rotational bands.  If they are shifted in such a way,
that they meet the highest level of the rotational band, which
occurs for $S=N s, E=N j s^2$, the resulting approximation is
very close to the true boundaries of the spectrum, compare the
thin curves in figures \xref{F-1}, \xref{F-2}, and \xref{F-3}.
Moreover, it turns out that the shifted parabolas exactly meet
the minimal and maximal energies of $S=N s - 1$
%--------------------------------------------------------
\begin{eqnarray}\label{26}
E_{\text{min}}
(N s -1)
&=&
j N s^2 + 2 (j_{\text{min}}-j) s
\\
E_{\text{max}}
(N s -1)
&=&
j N s^2 + 2 (j_{\text{max}}-j) s
\ .
\nonumber
\end{eqnarray}
%--------------------------------------------------------
Identifying the curvature for the lower parabola with
$(j-j_{\text{min}})/N$ provides a very useful
approximation for the curvature of the true low-lying rotational
band, and thus complements the derivations of \cite{ScL:PRB}
which are based on the sublattice structure of the spin array.
The classical counterparts $E_{\text{lower}}(S)$ and
$E_{\text{upper}}(S)$ of the bounding parabolas are also rather
close to the true boundaries of the spectrum.

By using these proposed methods one can predict the detailed
shape of the spectrum and related properties of molecular
magnets without diagonalizing a huge Hamilton matrix.

The overwhelming numerical evidence for the existence of
a rotaional band of minimal energies \cite{ScL:PRB} 
%--------------------------------------------------------
\begin{eqnarray}\label{27}
E_{\text{min}}(S,s)
\approx
\frac{D(s)}{N} S (S+1)
+
E_0(s)
\end{eqnarray}
%--------------------------------------------------------
for general Heisenberg spin systems cannot be mere coincidence.
We believe that the derived bounding parabolas provide a first
step towards proving this general behaviour.

\acknowledgments
We thank the National Science Foundation and
the Deutscher Akademischer Austauschdienst for supporting a
mutual exchange program.  The Ames Laboratory is operated for
the United States Department of Energy by Iowa State University
under Contract No. W-7405-Eng-82.

\end{document}